\documentclass[12pt,thmsa]{article}

\title {Quantum observables, Lie algebra homology and TQFT.}
\author {Albert Schwarz\thanks {Research supported in part by NSF grant No.
DMS-9801009, by IHES and by Mittag-Leffler Institute }\\
Math. Dept., University of California, Davis, CA 95616, USA\\
schwarz@math.ucdavis.edu}

\begin{document}
\maketitle
\begin {abstract}
  
  Let us consider a Lie (super)algebra $G$ spanned by $T_{\alpha}$ where
$T_{\alpha}$ are quantum observables in BV-formalism. It is proved that for
every tensor $c^{\alpha _1...\alpha _k}$ that determines a homology class
of the Lie algebra $G$ the expression $c^{\alpha_1...\alpha _k}T_{\alpha
_1}...T_{\alpha _k}$ is again a quantum observables. This theorem is used
to construct quantum observables in BV sigma-model. We apply this
construction to explain Kontsevich's results about the relation between
homology of the Lie algebra of
Hamiltonian vector fields and topological invariants of manifolds.
 \end {abstract}

  In this paper we obtain some general results about quantum observables in
Batalin-Vilkovisky formalism. Namely, we formulate conditions when a
polynomial expression in quantum variables is again a quantum variable.
This
result can be applied in many concrete situations; we use it to analyze
topological BV sigma-model introduced in [4]. We consider BV sigma-model
where the target is a linear symplectic space; the model can be considered
as BV formulation of abelian Chern-Simons theory. It is well known [6]
that the partition function of this theory is related to Ray-Singer
torsion. We show that appropriate correlation functions are related to
topological invariants considered in [5].

  First of all we will remind the notions of cohomology and homology of Lie
algebras and describe the relation of these notions to observables in BV
formalism.

  Let $G$ denote a Lie algebra with generators $T_{\alpha }$. To define the
cohomology of $G$ we consider the differential
 $Q={1\over 2}f_{\beta \gamma }^{\alpha }c^{\beta }c^{\gamma }{\partial
\over \partial c^{\gamma }}$
acting on the space of polynomials depending on anticommuting ghost
variables $c^{\alpha }$. We have $Q^{2}=0$ and therefore we can define
cohomology as Ker $Q/$Im $Q$. To define $k$-dimensional cohomology
we restrict ourselves to the space of polynomials of degree $k$, (or,
equivalently to antisymmetric tensors with $k$ lower indices). Homology of
the Lie algebra $G$\ can be defined as a linear space, dual to the
cohomology; one can give direct definition in terms of a differential
acting on antisymmetric tensors with upper indices or on functions
depending on anticommuting variables $c_{\alpha }$. If the Lie algebra
$G$ has dimension $N$ and $N$-dimensional (co)homology of $G$ 
is non-trivial, there exists an isomorphism between $k$-dimensional homology
and $(N-k)$-dimensional cohomology of $G$ (duality). This isomorphism can be
constructed by means of duality of tensors, or, equivalently, by means of
superanalog of the Fourier transform: if $f(c^{1},...,c^{N})$ is a cocycle,
then the function
\begin {equation}
\int f(c^{1},...,c^{N})\exp (c^{\alpha }c_{\alpha })dc^{1}...dc^{N}
\end {equation}
can be considered as a cycle of the Lie algebra $G$. (The volume element
on $\Pi G$ used in the definition of Fourier transform should correspond
to a nontrivial cohomology class of $G$.)  If $G$\
is an infinite-dimensional Lie algebra one should be more careful with the
definition of homology and cohomology; however, again with
appropriate definitions we obtain duality between $(\infty -k)$
-dimensional cohomology and $k$-dimensional homology.

  One can generalize the definition of homology and cohomology to the case
when $G$ is a Lie superalgebra. In this case one should understand the
antisymmetry of tensors that determine homology and cohomology classes in 
the sense of superalgebra. If we use ghost variables $c_{\alpha}$ and
$c^{\alpha}$ we should assume that $\varepsilon (c_{\alpha})=\varepsilon
(c^{\alpha})=\varepsilon (T_{\alpha})+1$ (i.e.the parity of these variables
is 
opposite to the parity of generators $T_{\alpha}$ of Lie superalgebra $G$.)
In other words, $Q$ acts on functions defined on $\Pi G$ in the case of
cohomology and on $\Pi G^*$ in the case of homology; $\Pi$ stand for
the parity reversion.

  We will work in the framework of BV-formalism (see [1], [2]).

  Our starting point is an $SP$-manifold $M$ (a manifold pasted together
from
domains in $R^{n|n}$ by means of transformations preserving volume and odd
symplectic form $dx^id\xi _i$). One can define an operator $\Delta $,
satisfying $\Delta ^2=0$, on an $SP$-manifold (in local coordinates $\Delta 
=\partial ^2/ \partial x^i\partial \xi_i$).
The operator $\Delta $ transforms a function $F$ into a function $\Delta
F={1\over 2}$
div $K_F$, where $K_F$\ stands for the Hamiltonian vector field,
corresponding to $F$ .
It is related to the odd Poisson bracket $\{ ,\}$ coresponding to the odd 
symplectic form $dx^id\xi _i$ by the formula
\begin {equation}
\Delta (KL)=\Delta K\cdot L+(-1)^{\varepsilon (K)}K\cdot \Delta
L+(-1)^{\varepsilon (K)}\{ K,L\}.
\end {equation}
  
  One can describe $SP$-manifold as an odd symplectic manifold
($P$-manifold) equipped with a volume element, that is compatible with
symplectic structure. (The compatibility condition is $\Delta ^2=0$ where
$\Delta F={1\over 2}$ div $K_F$.) 

  An action functional in the BV-formalism
is an even function $S$\ on an $SP$-manifold $M$, obeying the
quantum master equation $\Delta$ exp${S\over \hbar}=0$ or equivalently $2
\hbar \Delta S+\{S,S\}=0$. A quantum observable is a function $A$ on $M$
satisfying $\hbar \Delta A+\{ A,S\} =0$, or $(\hbar \Delta
+{\hat {Q}})A$, where ${\hat {Q}}A=\{A,S\}$. It is important to notice,
that a quantum observable $A$ in our definition is not necessarily an even
function on $M$. The expression
$$\int_{L}Ae^{{S\over \hbar }}d\lambda $$
where $L$ is a Lagrangian submanifold of $M$ depends only on homology
class of $L$ and has the meaning of expectation value of $A$. If a quantum
observable $A$ can be represented in the form $A=\hbar \Delta B+\{ B,S\}$ 
then its expectation value vanishes; we say in this case that the
observable $A$ is trivial.

Proposition. a) If $A$ and $B$ are quantum observables then $\{ A,B\}$ is
also a quantum observable. (In other words quantum observables constitute a
Lie superalgebra.)

  b)Let us suppose that quantum observables $T_{\alpha}$ span a Lie
(super)algebra $G$ (i.e. $\{ T_{\alpha},T_{\beta}\}
=f^{\gamma}_{\alpha\beta}T_{\gamma}$) and that antisymmetric tensor
$c^{\alpha _1,...,\alpha _k}$ specifies a homology class of $G$. Then
$$T=c^{\alpha _1...\alpha _k}T_{\alpha _1}...T_{\alpha _k}.$$
is also a quantum observable. If the tensor $c^{\alpha _1...\alpha _k}$
belongs to the trivial homology class the corresponding observable is also
trivial.

  One can derive these statements by means of straightforward calculations
based on the definition of homology, on the relation
 $\{S,KL\}=\{S,K\}\cdot L+(-1)^{\varepsilon (K)}K\cdot \{S,L\}$ and
on the formula (2).

  Notice, that it was proved in [3] that with appropriate definition of
symmetries in BV-formalism one can construct a one-to-one correspondence
between quantum observables and symmetry transformations. The statement a)
follows immediately from this result (see Appendix for more details).

There exist serious difficulties in application of the above statement to
quantum field theory. In quantum field theory we should consider
infinite-dimensonal manifolds; the operator $\Delta$ is ill-defined and
therefore we encounter the standard problems related to ultraviolet
divergences in the definition of quantum observable. However, we will show
that formal application of our theorem leads to interesting results.
 
  Let us apply the above consideration to the BV sigma model [4]. Let us
denote by $E$ the space of maps of $\Sigma =\Pi TX$\ into $M$. If $X$ is an
odd-dimensional compact manifold and $M$ is a manifold, equipped with even
symplectic structure we define an odd symplectic structure on $E$ by means
of the form
$$\sigma _f(\varphi _1,\varphi _2)=\int_{\Sigma }\sigma _{f(\gamma)}
(\varphi _1(\gamma ),\varphi _2(\gamma ))d\mu$$
where $\varphi _1$ and $\varphi _2$ are elements of the tangent space $T_{f}
(E)$ (i.e. infinitesimal variations of $f$), $d\mu $ stands for the
standard measure on $\Sigma =\Pi TX,\gamma =(x^{\alpha },\xi ^{\alpha })$,
where $x^{\alpha }$ are coordinates on $X$\ and $\xi ^{\alpha }$ can be
identified with anticommuting differentials of coordinates, $\sigma =\sigma
_{ab}dz^{a}dz^{b}$ denotes the symplectic
form on $M$. The de Rham differential induces an odd vector field $Q$ on $E$ 
obeying $\{Q,Q\} =0$ and preserving the odd symplectic structure on $E$. The
Hamiltonian $S$ of this vector field satisfies $\{S,S\} =0$; by definition
$S$ is an action functional of BV sigma model. In other words, $S$ can be
determined from the equation
$${\delta S\over \delta \varphi ^a}=\sigma _{ab}(\varphi )\xi ^{\alpha}
{\partial \varphi ^b\over \partial x^{\alpha }}$$
It is important to emphasize the obvious fact that the action functional $S$
is invariant with respect to symplectic transformations of $M$ and to
diffeomorphisms of $X$. This means that the corresponding physical
quantities should be invariants of smooth structure of $X$ and
symplectic structure of $M$ (if the symmetries are not anomalous).

  Notice that $S$ satisfies not only classical, but also quantum master
equation (i.e. $\Delta S=0$). The space $E$ is infinite-dimensional,
therefore $\Delta $ is ill-defined; however we can use the fact that the
equation $\Delta S=0$ is equivalent to the equation $divQ=0$. In other
words, we should check that the vector field $Q$ is volume preserving. It
is clear that for reasonable definition of the ''volume element'' on $E$ 
this fact is correct.

  This construction can be modified in different ways. In particular, one
can
assume that $M$ is a symplectic $Q$-manifold (i.e. it is equipped with an
odd vector field $Q$, obeying $\{Q,Q\} =0$ and preserving symplectic
structure.) One can suppose, that $X$ is even-dimensional; then instead of
even symplectic structure on $M$ one should consider an odd symplectic
structure. In all these cases one can construct an action functional of BV 
sigma model. It is shown in [4], that many interesting models (including
Chern-Simons theory) can be regarded as particular cases of general
construction described above.

  We will consider the simplest case when $M$ is a linear symplectic space
and $X$ is odd-dimensional, 
however, our main ideas can be applied also in general situation. If $M$ is
linear, we can assume that $\sigma _{ab}$ does not depend on $\varphi $ and
represent the action functional in the form
\begin {equation}
S=\int \sigma _{ab}\varphi ^a\xi ^{\alpha }{\partial \varphi ^b\over 
\partial x^{\alpha }}.
\end {equation}
(Indices $a,b$ run over the set $\{ 1,...,\dim M\} $.)
 
  One can consider the fields $\varphi ^a$ as inhomogeneous differential
forms on $X$; then the functional (3) can be rewritten as follows:
$$S=\int \sigma _{ab}\varphi ^ad\varphi ^b.$$ We see that corresponding
field theory can be interpreted as abelian Chern-Simons theory in
BV-formalism. We will study correlation function of this theory. (The
partition function can be expressed in terms of Ray-Singer torsion [6].)

  Let us consider the Lie algebra $H$ of polynomial Hamiltonian vector
fields on $M$. It can be identified with the Lie algebra of polynomials $
P(\varphi)$\ on $M$ with respect to Poisson bracket. (More precisely,
we should factorize the algebra of polynomials with respect to constants.)
One can consider $H$ as an algebra of symmetries of (3). Namely, every
polynomial $P(\varphi)$ determines a vector field $\delta \varphi =\sigma
^{ab}{\partial P\over \partial \varphi ^b}$ on $M$; the same formula
where $\varphi ^a$ are regarded as functions on $\Sigma $ specifies a
vector field on $E$, preserving the functional $S$. Corresponding
Hamiltonian has the form
$$\Psi _P=\int P(\varphi (x,\xi ))dxd\xi .$$
It is clear that $\{\Psi _P,S\} =0$, i.e. $\Psi _P$ is a classical
observable.

  To check that $\Psi _P$ is a quantum observable, one should prove that
corresponding vector field  preserves ''volume element'' on $E$ (i.e. there
exists no quantum anomaly). In one-loop approximation one can express
quantum anomaly in terms of Seeley coefficients and zero modes [6]. We
assumed 
that the manifold $X$ is odd-dimensional; it follows from this
assumption that relevant Seeley coefficients vanish; requiring that $X$ be
a homology sphere we arrive at a conclusion that the symmetries at hand are
not anomalous (see comments at the end of the paper for more precise
statements).

  It follows from the above consideration that every homology class of the
Lie
algebra $H$ corresponds to a quantum observable. Kontsevich proved [5] that 
every homology class of the graph complex constructed in [5] generates a
homology class of $H$.
We will calculate the expectation value of corresponding quantum observable;
the expression we obtain is closely related to the constructions of [5].
Observables we study can be represented as $\sum c_{\Gamma }\Psi _{\Gamma}$. 
Here $\Gamma$ is an oriented graph, $\Psi _{\Gamma}$ can be constructed by
means of standard Feynman rules; every edge contributes $\sigma _{ab}$,
every vertex contributes $\varphi ^{a_1}(z)...\varphi ^{a_k}(z)$. (We
assign a point $z\in \Sigma $ to every vertex and two indices $a,b\in A$ to
every edge. $\Psi _{\Gamma }$ can be obtained by means of integration over
all points of $\Sigma $ and summation over all indices).

  One can check that in the case when $\sum c_{\Gamma }\Gamma $ is a cycle
in the
graph complex, constructed in [5], the expression $\sum c_{\Gamma }\Psi
_{\Gamma}$ is a quantum observable. To verify this fact we fix a basis of
$H$ consisting of generators $T^{a_1....a_k}$ that correspond to
Hamiltonians $\varphi ^{a_1}...\varphi ^{a_k}$. Using Feynman rules we
construct for every oriented graph $\Gamma $ an element $\Phi _{\Gamma }$
of antisymmetric part of product $H^{\otimes l}$, where $l$ is the number of
vertices of $\Gamma $. Namely we assign $\sigma _{ab}$ to every edge and $
T^{a_1....a_k}$ to every vertex; to define $\Phi _{\Gamma }$ we take
tensor product of elements of $H$, corresponding to every vertex, multiplied
by the product of all $\sigma _{ab}$, corresponding to the vertices, and
sum
over repeating indices. In the case when $\sum c_{\Gamma }\Gamma $ is a
cycle in the graph complex, one can prove that $\sum c_{\Gamma }\Phi
_{\Gamma}$ determines an element of the homology group of the Lie algebra
$H$ [5]. This remark together with the relation between homology of Lie
algebras and observables leads to the consideration of the observables
$\sum
c_{\Gamma }\Psi _{\Gamma }$. We should expect that expectation values of
these observables are invariants of smooth manifold $X$; we will see that
they coincide
with invariants constructed in [5].

  Now we can calculate expectation values of observables at hand using
again the Feynman diagram technique. (We consider only connected
diagrams; in other words we normalize the observables dividing by  the
partition finction.) We obtain new edges connecting the same
vertices, the propagator corresponding to new edges has the form $\sigma
^{ab}\omega (z_1,z_2)$. (The propagator depends on the choice of
Lagrangian submanifold in the functional integral for the expectation
value; see [5] for the description of $\omega (z_1,z_2)$.) We do not
discuss the derivation of the diagram technique in detail;
very similar problem is thoroughly analyzed in [7]. Taking into account
that $\sigma _{ab}\sigma ^{bc}=\delta ^{ac}$ we see that one can get non-zero
contribution only in the case when new edges coincide with old ones. The
contribution of a graph $\Gamma $ to the expectation value of $\sum
c_{\Gamma}\Psi _{\Gamma}$ is equal to $c_{\Gamma }$, multiplied by the
integral over $\Sigma$ of the product of factors $\omega (z_i,z_j)$,
corresponding to the edges of $\Gamma $ (here $z_i,z_j\in \Sigma $).
We see that the diagrams for expectation values of observables $\sum
c_{\Gamma}\Psi_{\Gamma}$ coincide with the diagrams for invariants of
smooth manifold $X$ defined in [5]. Our consideration can be regarded as a
heuristic derivation of Kontsevich invariants. 

We skipped several subtle points in our exposition. First of all the
assumption that our manifold $X$ is a homology sphere is not
sufficient to exclude anomalies alltogether. However, the anomaly
related to non-zero homology groups in dimensions 0 and $\dim X$ can be
easily taken into account. ( Kontsevich removes one point from $X$ and
considers the result as a  euclidean space with topology perturbed in a
compact subset to avoid this anomaly.)  Our consideration prompts an idea
how to generalize  Kontsevich constructions to the case when $X$ is not
a homology sphere. We mentioned already that the partition function of the
model we used is related to Ray-Singer torsion. It is well known that
in the case when the manifold $X$ is not acyclic one should consider the
torsion as a measure on the linear superspace $ H(X) $ ( on the direct sum
of cohomology groups of $X$ ); this measure is an invariant of $X$ ( see
for example [6], [8]). The expectation values of observables
we
studied can lead to invariants of similar nature.

It seems that the above consideration can be used also to generalize
Kontsevich invariants to the case of manifolds with boundary.

{\bf Acknowledgment.} I am deeply indebted to M.Kontsevich for many useful
discussions.

\centerline {{\bf Appendix.  Quantum and classical observables.}}
 
  Let us formulate some of results of [3]. We consider an SP-manifold $M$
with volume element $d\mu$. New volume element $d{\tilde
{\mu}}=e^{\sigma}d\mu$
determines an SP-structure on $M$ iff $\Delta\sigma+{1\over 4}\{ \sigma
,\sigma \}=0$. The functional ${\tilde {S}}=S-{1\over 2\hbar}\sigma $
satisfies
the quantum master equation with respect to the new SP-structure and is
physically equivalent to the original action functional $S$, i.e. for every
observable $A$ we nave 
$$\int _L Ae^{-{\tilde {S}}/\hbar}d{\tilde {\lambda}}=\int _L
Ae^{-S/\hbar }d\lambda. $$
(An observable with respect to $S$ can be considered also as an observable
with respect to ${\tilde {S}}$.) One can use the freedom in the choice of
volume element on $M$ to replace the action functional $S$ with physically
equivalent action functional ${\tilde {S}}=0$; then an observable can be
characterized as a function $A$ obeying ${\tilde {\Delta}}A=0$ or as a
vector field preserving the odd symplectic structure and the new volume
element. This description of observables makes the statement a) of
Proposition 1 obvious and simplifies the proof of the statement b). 
 
  It follows from the results of [3] that in quantum BV-formalism one can
consider an action functional as a volume element on an odd symplectic
manifold, that is compatible with symplectic structure (determines an
SP-structure). However, such an approach is not convenient if we would like
to relate quantum BV-formalism to the classical one. 

  In classical case the main object is a solution to the classical master
equation $\{ S,S\} =0$ where $S$ is an even function on odd symplectic
manifold. A classical observable is a function $A$ obeying $\{ A,S\} =0$.
It is clear the functional $S$ is invariant with respect to the Hamiltonian
vector field $K_A$ corresponding to the observable $A$. We see that in
classical case the identification of observables and symmetries and the
fact that observables constitute a (super) Lie algebra are obvious. We can
arrive at these notions taking the limit $\hbar \rightarrow 0$ in quantum
master
equation and in the definition of quantum observable. The most important
action functionals and observables obey both quantum and classical
equations (i.e. satisfy $\Delta S=\{ S,S\} =0,\ \ \ \Delta A=0,\ \ \  \{
A,S\} =0$). Let us consider a solution $S$ to the classical master equation
$\{ S,S\} =0$ and a set of observables $T_{\alpha}$ that constitute a Lie
algebra $G$.

  In other words, we suppose that we have functionals $T_{\alpha }$\ on $M$
obeying $\{S,T$\ $_{\alpha }\}=0$ and $\{T_{\alpha },T_{\beta }\}=
f_{\alpha \beta }^{\gamma }T_{\gamma}$, where $f_{\alpha \beta }^{\gamma}$ 
are structure constants of the Lie algebra $G$. Then we can consider 
a new odd symplectic manifold ${\tilde {M}}$
adding ghosts $c^{\alpha }$ and corresponding antifields $c^*_{\alpha}$. 
(More formally, ${\tilde {M}}=M\times \Pi G\times G^*$.) One can extend
$S$ to ${\tilde {M}}$, taking
$${\tilde {S}}=S+{1\over 2}f_{\alpha
\beta}^{\gamma}c^{\alpha}c^{\beta}c_{\gamma}^{*}$$
We obtain a new solution to the master equation. We will be interested in
classical observables corresponding to the functional ${\tilde {S}}$, i.e.
functions 
on ${\tilde {M}}$, obeying $\{ {\tilde {S}},T\} =0.$ It
is easy to check that observables depending only on the ghosts $c^{\alpha }$
can be identified with cohomology of the Lie algebra $G$. 

  If the action functional $S$ satisfies both classical and quantum master
equation (i.e. $\{ S,S\} =0$ and $\Delta S=0$) and $T_{\alpha}$ are not
only classical, but also quantum observables, one can prove the following
relation:
\begin {equation}
\int _L Ae^{S/\hbar }d\lambda =\int _{\tilde {L}}{\tilde {A}}e^{-{\tilde
{S}}/\hbar }d{\tilde {\lambda}}.
\end {equation}
Here $A$ is an observable for $S$, corresponding to a homology 
$\alpha$ of the algebra $G$ and ${\tilde {A}}$ is an
observable for ${\tilde {S}}$ corresponding the cohomology class of $G$
that is dual to $\alpha$. (We define a Lagrangian submanifold ${\tilde
{L}}\subset {\tilde {M}}$ as a direct product of $L$ and Lagrangian
submanifold in $\Pi G\times G^*$ that is singled out by the equations
$c^*_{\alpha}=0$). To define duality between Lie algebra homology
and cohomology we need a non-trivial class of $N$-dimensional
(co)homology of $G$; the same class should be used to define a volume
element in ${\tilde {M}} =M\times \Pi G\times G^*$. The formula (4) follows
immediately from the definition of duality in terms of Fourier transform.)

\vskip .1in
\centerline {{\bf References.}}
\vskip 1in

  1. I.Batalin, G.Vilkovisky, Quantization of gauge theeories with linearly
dependent generators. Phys, Rev. D 28 (1983), 2567-2582.

  2. A. Schwarz, Geometry of Batalin-Vilkovisky quantisation, Comm. Math.
Phys. 155 (1993), no. 2, 249-260.

  3. A.Schwarz, Symmetry transformations in Batalin-Vilkovisky formalism,
Lett. Math. Phys. 31 (1994), no. 4, 299-301.

  4. M.Alexandrov, A.Schwarz, O,Zaboronsky, M.Kontsevich, The geometry of
the master equation and topological quantum field theory, Internat. J.
Modern Phys. A 12 (1997), no. 7, 1405-1429.

  5. M.Kontsevich, Feynman diagrams and low-dimensional topology, First
European Congress of Mathematics, Vol.II (Paris,1992), 97-121, Progr.
Math., 120, Birkhaeuser, Basel, 1994.

  6. A.Schwarz, The partition function of a degenerate functional, Comm.
Math. Phys. 67 (1979), 1-16.

  7. S.Axelrod, I.Singer, Chern-Simons perturbation theory, in:
Differential geometric methods in theoretical physis, vol. 1, pp. 3-45, New
York, 1991.

  8, A. Schwarz, Yu. Tyupkin, Quantization of antisymmetric tensors
and Ray-Singer torsion, Nucl. Phys. B 242 (1984) 436-446.

 \end{document}